\pdfoutput=1
\documentclass[lettersize,journal]{IEEEtran}
\usepackage{amsmath,amssymb,amsfonts,bm}
\usepackage{graphicx}
\usepackage{booktabs}
\usepackage{array}
\usepackage{cite}
\usepackage{url}
\usepackage{balance}
\usepackage{xcolor}
\usepackage{tikz}
\usetikzlibrary{arrows.meta,positioning,calc}
\graphicspath{{figures/}}

\definecolor{mlblue}{HTML}{0072BD}
\definecolor{mlorange}{HTML}{D95319}

\newtheorem{theorem}{Theorem}
\newtheorem{lemma}{Lemma}
\newtheorem{proposition}{Proposition}
\newtheorem{corollary}{Corollary}
\newtheorem{remark}{Remark}

\newcommand{\None}{N_1}
\newcommand{\Nstar}{N_1^{\star}}
\newcommand{\NR}{N_{\mathrm{R}}}
\newcommand{\KD}{K_{\mathrm{D}}}
\newcommand{\Ts}{T_{\mathrm{s}}}
\newcommand{\Ds}{D_{\mathrm{L}}}
\newcommand{\tauS}{\tau}
\newcommand{\Pe}{P_{\mathrm{e}}}
\newcommand{\Ptr}{P_{\mathrm{tr}}}
\newcommand{\Prb}{\mathbb{P}}
\newcommand{\E}{\mathbb{E}}
\newcommand{\Bin}{\mathrm{Bin}}
\newcommand{\erfop}{\operatorname{erf}}
\newcommand{\Sp}{S_{\mathrm p}}
\newcommand{\dd}{\,\mathrm{d}}
\newcommand{\um}{\mu\text{m}}
\newcommand{\pA}{p_{\mathrm{A}}}

\makeatletter
\g@addto@macro\normalsize{\setlength{\abovedisplayskip}{4pt plus 2pt minus 1pt}\setlength{\belowdisplayskip}{4pt plus 2pt minus 1pt}\setlength{\abovedisplayshortskip}{2pt plus 1pt}\setlength{\belowdisplayshortskip}{2pt plus 1pt}}
\makeatother

\begin{document}
\bstctlcite{IEEEexample:BSTcontrol}

\title{A Closed-Form Molecule-Release Rule for Diffusion-Based Molecular Communications with Ligand Receptors}

\author{Eren~Kural,~\IEEEmembership{Student Member,~IEEE}, and Murat~Kuscu,~\IEEEmembership{Member,~IEEE}\thanks{The authors are with the Nano/Bio/Physical Information and Communications Laboratory (CALICO Lab), and the Center for neXt-generation Communications, Department of Electrical and Electronics Engineering, Ko\c{c} University, Istanbul 34450, T\"urkiye (e-mail: \{ekural24, mkuscu\}@ku.edu.tr).}\thanks{This work was supported by the Scientific and Technological Research Council of T\"urkiye (T\"UB\.ITAK) under Grant 123E516.}}

\maketitle

\begin{abstract}
The number of molecules released per bit is a fundamental design variable of diffusion-based molecular communication
(MC), and ligand-receptor reception breaks the more-is-better intuition. Too few molecules leave the bound-receptor observations buried in binding noise, while too many amplify the accumulated intersymbol interference and saturate the finite receptor population, again making the observations indistinguishable.  Reliability therefore peaks in an interior operating region whose location seems to require an exhaustive search over the channel dynamics. In this paper, we show that this search can be obviated for a biologically plausible receiver that compares consecutive bound-receptor counts without channel state information or a decision threshold. We derive a closed-form transmission rule, which sets the number of molecules released per bit such that the receptor dissociation constant equals the geometric mean of the two bit-conditioned received concentration levels, prove that it exactly minimizes the bit error probability of a memoryless binomial receptor model, and express it in the physical channel parameters through an Euler--Maclaurin evaluation of the interference. Time-domain Monte Carlo sweeps of the channel and receptor parameters, corroborated by particle-based simulations, show that the empirically optimal release count coincides with the prediction or lies above it by a small factor.
\end{abstract}


\section{Introduction}
\label{sec:intro}

\IEEEPARstart{W}{hen} engineered devices approach the scale of living cells, conventional wireless communication becomes impractical. Cells communicate chemically by releasing molecules that diffuse and bind to receptor proteins. Molecular communication (MC) adopts this strategy as an engineering paradigm, encoding information in the type, number, or timing of released molecules, and is a key enabler of the Internet of Bio-Nano Things~\cite{Akan2017Fundamentals,Kuscu2019ArchitecturesSurvey,Jamali2019ChannelTutorial}.

The apparent simplicity of the diffusion-based MC channel conceals a physics of unusual character. A molecule released by a transmitter performs a random walk, and the concentration observed at a receiver therefore rises to a peak and then decays with the heavy $t^{-3/2}$ tail of three-dimensional diffusion. Molecules released for one symbol linger through many subsequent symbol intervals, and the resulting intersymbol interference (ISI) intensifies with the symbol rate~\cite{Jamali2019ChannelTutorial,Tepekule2015ISI}. Reception adds its own complications. A practical MC receiver, whether a living cell or an engineered device built on affinity-based biosensors~\cite{Kuscu2016PhysicalDesign,Kuscu2021Graphene}, does not measure concentration directly but senses it through a finite population of receptors whose binding and unbinding are stochastic~\cite{Berg1977Chemoreception,Pierobon2011LigandNoise}. At equilibrium, the probability that a receptor is occupied follows the Langmuir law $q(C)=C/(C+\KD)$, where $C$ is the local ligand concentration and $\KD$ the dissociation constant of the receptor, a curve that is steep only within about a decade of $\KD$. Far below this range hardly any receptors bind and the signal is lost in binding noise, and far above it the response saturates and erases the difference between symbols~\cite{Ahmadzadeh2016ReactiveReceiver,Lotter2021SaturatingReceiver,Kuscu2019ChannelSensing}.

Together, these two features give an elementary design decision a nonobvious answer, namely the number of molecules $\None$ that an on-off keying (OOK) transmitter should release to convey bit-1. The intuition inherited from additive-noise channels, that more transmit power, here more released molecules, yields higher reliability, pervades much of MC system design, and with receptor-based reception under ISI it fails. Too few molecules leave both bit-conditioned occupancies near zero and masked by binding fluctuations, while too many raise the residual concentration from past bit-1 symbols in proportion and push the receptor population into saturation, where the occupancies converge again near one. Reliability is therefore not monotone in the release count, and there is an interior operating region, a communication-theoretic analogue of the therapeutic window in pharmacology. Locating this window also matters in practice. Molecules must be synthesized or stored at a real energetic cost~\cite{Kuran2010Energy}, excessive release pollutes the medium shared with other channels, and a bioactive payload turns over-release into over-dosing~\cite{Chahibi2013DrugDelivery}.

This window depends on receptor affinity and number, the distance, the diffusivity, the symbol duration, and the interference history, and finding it seems to require exhaustive simulation or trial and error, neither suited to a resource-limited bio-nano device. This paper shows that no search is needed for the comparator-type receiver introduced next, whose optimal release count admits a closed form.

The receiver we consider decides bit-1 whenever the current bound-receptor count exceeds the previous one. Such consecutive-sample comparator detection requires no channel state information (CSI) and no decision threshold, both costly to acquire and maintain at this scale, and it mirrors the temporal comparison of receptor occupancy by which bacteria navigate chemical gradients~\cite{Segall1986TemporalComparisons}. Detectors of this family have been studied in MC as adaptive-threshold and noncoherent schemes robust to channel uncertainty~\cite{Damrath2016ATD,Li2016LocalConvexity}. With the receiver stripped of every adjustable parameter, the release count is the only remaining degree of freedom.

\subsection{Related Work}
\label{sec:related}
Detection in diffusion-based MC spans optimal sequence detection and equalization~\cite{Kilinc2013ReceiverDesign,Noel2014OptimalReceiver}, low-complexity threshold schemes~\cite{Llatser2013Detection,Damrath2016ATD}, and ISI-mitigating modulation and filtering~\cite{Tepekule2015ISI}. The ligand-receptor front end has been modeled with increasing fidelity, covering binding noise~\cite{Pierobon2011LigandNoise}, reactive and reversible receiver surfaces~\cite{Deng2015ReversibleAdsorption,Ahmadzadeh2016ReactiveReceiver}, receptor-based maximum-likelihood detection and channel sensing~\cite{Kuscu2018MLReceptors,Kuscu2019ChannelSensing,Kuscu2022Interference}, and the saturation and memory effects of finite receptor populations~\cite{Lotter2021SaturatingReceiver,Zheng2026LangmuirChannel}. Optimum detection for concentration-encoded signals with reception noise~\cite{Mahfuz2014SamplingDetection} and fully chemical or learning-based receivers~\cite{Heinlein2025ChemicalReceiver,Bilge2026ReservoirDetection} have also been pursued.

On the transmitter side, the released amount has long been recognized as a design variable. Early work controlled the transmission rate to match the reaction capacity of chemically reacting receivers~\cite{Nakano2013RateControl}. Later studies optimized the released amount for reliability, through symbol-by-symbol adaptation to the ISI seen by a fixed-threshold receiver~\cite{Movahednasab2016AdaptiveTx}, power control scaled with the estimated distance and residual channel content in mobile settings~\cite{Jing2021PowerControl}, joint optimization of release count and detection threshold for mobile drift-diffusion channels~\cite{Chouhan2019OptimalMolecules}, MIMO release vectors~\cite{Cheng2023OptimalMIMO}, allocation of a limited molecule or energy budget across transmissions~\cite{Musa2023EnergyHarvesting,Jing2025EnergyAllocation}, and molecule-budget versus error-rate trade-offs~\cite{Jing2025MoleculeBERTradeoff}. In these later studies, the receiver counts the molecules inside a transparent or absorbing volume, an abstraction that omits the saturable binding process through which biological and biosensor-based receivers actually observe the channel~\cite{Kuscu2016PhysicalDesign}. Moreover, their optimal release count emerges from a numerical optimization tied to a particular threshold detector, without a closed form. Closest to this study is our previous work~\cite{Kuscu2026AdaptiveReceiver}, which showed that a receiver able to tune its receptor affinity performs best when $\KD$ matches the geometric mean of the two received levels. This paper solves the inverse and arguably more practical problem, since receptor affinity is fixed by the deployed proteins whereas regulating a release count is exactly what MC transmitter architectures are designed to do~\cite{Kuscu2019ArchitecturesSurvey}. No prior work provides a closed-form release count for receptor-based reception under ISI.

\subsection{Contributions}
\label{sec:contrib}
Our contributions are fourfold. First, we formulate the transmission problem, namely the choice of the release count $\None$, for OOK over a diffusion-based channel with Langmuir receptors and a CSI-free consecutive-sample comparator, and prove that the bit error probability of the associated memoryless binomial model is minimized by the closed-form release count $\Nstar=\KD/\sqrt{\alpha\beta}$ (Theorem~\ref{thm:main}), where $\alpha\None$ and $\beta\None$ are the mean concentrations observed at the sampling instant for a current bit of 0 and 1, $\alpha$ set by the accumulated ISI alone and $\beta$ by the ISI plus the current pulse (Section~\ref{sec:model}). Second, we show that as a function of $\None/\Nstar$ the model error probability depends on the channel only through the ratio $\beta/\alpha$ of the two levels, the ISI ratio, while $\Nstar$ itself scales linearly with $\KD$, hence even a twofold error in either received-level coefficient displaces the operating point by only a factor of $\sqrt2$. Third, we make the rule explicit in the physical channel parameters through a closed-form expression for the accumulated ISI that combines the exact leading interference terms with an Euler--Maclaurin evaluation of the slowly decaying diffusion tail, together with a fully elementary variant. Fourth, we validate the rule against the reaction-diffusion physics that the analysis abstracts away, with time-domain Monte Carlo sweeps of the receptor number, affinity, diffusion coefficient, and distance against the symbol duration and particle-based Smoldyn simulations of selected conditions. Wherever detection is possible at all, the empirically optimal release count coincides with the prediction or lies within a small factor above it.

The remainder of the paper is organized as follows. Section~\ref{sec:model} presents the system model, Section~\ref{sec:rule} derives the closed-form molecule-release rule and its properties, Section~\ref{sec:methods} describes the simulation framework, Section~\ref{sec:results} presents the simulation results, Section~\ref{sec:discussion} discusses practical implications and limitations, and Section~\ref{sec:conclusion} concludes the paper.

\section{System Model}
\label{sec:model}

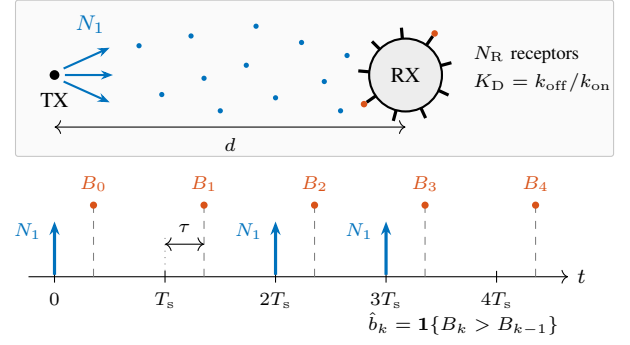
\begin{figure}[!t]
\centering
\begin{tikzpicture}[scale=0.86, every node/.style={font=\footnotesize}]
  \draw[rounded corners=1pt, fill=gray!4, draw=gray!45] (-0.6,-1.25) rectangle (8.6,1.15);
  \fill[black] (0,0) circle (0.075);
  \node[below=2pt] at (0,-0.05) {TX};
  \draw[-{Stealth[length=1.6mm]},thick,mlblue] (0.14,0.10) -- (0.85,0.42);
  \draw[-{Stealth[length=1.6mm]},thick,mlblue] (0.16,0) -- (0.9,0);
  \draw[-{Stealth[length=1.6mm]},thick,mlblue] (0.14,-0.10) -- (0.85,-0.42);
  \node[mlblue,anchor=south] at (0.55,0.5) {$\None$};
  \foreach \x/\y in {1.3/0.45,1.65/-0.3,2.1/0.6,2.55/-0.55,2.95/0.15,3.4/-0.35,3.8/0.5,4.15/-0.1,4.5/0.3,4.4/-0.55,3.1/0.75,2.3/-0.05}
     \fill[mlblue] (\x,\y) circle (0.04);
  \draw[fill=gray!14,thick] (5.4,0) circle (0.56);
  \node at (5.4,0) {RX};
  \foreach \a in {15,55,95,135,175,215,255,295,335}
     \draw[very thick] (5.4,0) ++(\a:0.56) -- ++(\a:0.17);
  \fill[mlorange] (5.4,0) ++(55:0.78) circle (0.05);
  \fill[mlorange] (5.4,0) ++(215:0.78) circle (0.05);
  \node[anchor=west,font=\scriptsize] at (6.3,0.32) {$\NR$ receptors};
  \node[anchor=west,font=\scriptsize] at (6.3,-0.12) {$\KD=k_{\mathrm{off}}/k_{\mathrm{on}}$};
  \draw[<->] (0,-0.8) -- (5.4,-0.8) node[midway,below=0.5pt,font=\scriptsize] {$d$};
  \begin{scope}[yshift=-3.1cm]
    \draw[->] (-0.4,0) -- (7.9,0) node[right] {$t$};
    \foreach \k/\b/\lab in {0/1/0,1/0/\Ts,2/1/2\Ts,3/1/3\Ts,4/0/4\Ts} {
      \draw (\k*1.7,-0.09) -- (\k*1.7,0.09);
      \node[below,font=\scriptsize] at (\k*1.7,-0.1) {$\lab$};
      \ifnum\b=1
        \draw[-{Stealth[length=1.8mm]},very thick,mlblue] (\k*1.7,0.03) -- (\k*1.7,0.85);
        \node[mlblue,font=\scriptsize,anchor=east] at (\k*1.7-0.06,0.7) {$\None$};
      \fi
      \draw[dashed,gray] (\k*1.7+0.6,0) -- (\k*1.7+0.6,1.1);
      \fill[mlorange] (\k*1.7+0.6,1.1) circle (0.055);
      \node[mlorange,above,font=\scriptsize] at (\k*1.7+0.6,1.17) {$B_{\k}$};
    }
    \draw[dotted,gray] (1.7,0.09) -- (1.7,0.62);
    \draw[<->] (1.7,0.5) -- (2.3,0.5) node[midway,above=1pt,font=\scriptsize] {$\tauS$};
    \node[font=\scriptsize,anchor=east] at (7.9,-0.72) {$\hat b_k=\mathbf 1\{B_k>B_{k-1}\}$};
  \end{scope}
\end{tikzpicture}
\caption{The MC channel. Top, bit $b_k=1$ releases $\None$ ligands, which diffuse over distance $d$ to a receiver carrying $\NR$ surface receptors of dissociation constant $\KD$ (orange, ligand-occupied). Bottom, the receiver samples the bound-receptor count $B_k$ at $k\Ts+\tauS$ and decides $\hat b_k=\mathbf 1\{B_k>B_{k-1}\}$.}
\label{fig:system}
\end{figure}

Fig.~\ref{fig:system} depicts the diffusion-based MC channel considered. A point transmitter (TX) releases bursts of ligand molecules that diffuse to a receiver (RX) carrying a fixed population of identical, independent receptors. We reduce the physics to the simplest description exhibiting the three central impairments, ISI accumulation, binding noise, and receptor saturation, derive the rule exactly within it, and then test it against increasingly detailed simulations in Sections~\ref{sec:methods} and~\ref{sec:results}.

\subsection{Diffusion Channel and OOK Signaling}
\label{sec:channel}
The TX is a point source that releases all molecules of a symbol at once, and the RX a point observer at distance $d$ whose receptors sense the local ligand concentration without perturbing it. The medium is an unbounded three-dimensional fluid without flow, in which the ligands diffuse with coefficient $\Ds$ without enzymatic degradation. The concentration produced at the RX by a unit release is then the free-space Green's function $h(t)=(4\pi\Ds t)^{-3/2}\exp[-d^2/(4\Ds t)]$, $t>0$, which peaks at $t_{\mathrm{peak}}=d^2/(6\Ds)$ and then decays only algebraically, as $t^{-3/2}$. Its time integral $\int_0^\infty h\dd t=1/(4\pi\Ds d)$ is finite, yet more than $90\%$ of it accrues after the peak and more than $40\%$ after $10\,t_{\mathrm{peak}}$, hence past symbols keep interfering.

The TX employs OOK, the simplest and most widely studied MC modulation. At the start of the $k$th symbol interval, $t_k=k\Ts$, it releases $N_k=\None b_k$ molecules, where $\{b_k\}$ is an independent and identically distributed (i.i.d.) Bernoulli bit sequence with $\Prb\{b_k=1\}=p_1$ and $p_0=1-p_1$. Because diffusion is linear, the mean concentration at the RX superposes the responses of all past releases,
\begin{equation}
    r(t)=\sum_{j:\,t_j<t}N_j\,h(t-t_j).
    \label{eq:superposition}
\end{equation}
Throughout, only $\None$ is a design variable, and the rest is fixed.

\subsection{Sampled Concentration and Two-Level Representation}
\label{sec:sampling}
The RX forms one observation per symbol, at time $k\Ts+\tauS$, in our numerical work at the isolated-pulse peak $\tauS=t_{\mathrm{peak}}$. The analysis holds for any phase $0<\tauS<\Ts$, which places the observation of symbol $k$ before the release of symbol $k+1$. In the stationary regime,
\begin{equation}
    r_k(\tauS)=N_k h(\tauS)+\sum_{\ell=1}^{\infty}N_{k-\ell}\,h(\tauS+\ell\Ts).
    \label{eq:rk}
\end{equation}
The bit-averaged strength of the second term is captured by the \emph{asymptotic sampled ISI coefficient}
\begin{equation}
    \Sp(\tauS)\triangleq\sum_{\ell=1}^{\infty}h(\tauS+\ell\Ts).
    \label{eq:spast}
\end{equation}
Because $h(t)\sim t^{-3/2}$, the sum converges slowly, the terms beyond the $L$th contributing a fraction of order $L^{-1/2}$, and it represents the interference of an infinitely long transmission, an upper bound on the ISI of any finite one.

Averaging \eqref{eq:rk} over the i.i.d.\ past bits yields the two bit-conditioned mean levels and their slopes,
\begin{equation}
    \begin{aligned}
    C_0(\None)&=\alpha\None,\qquad C_1(\None)=\beta\None,\\
    \alpha&=p_1\Sp(\tauS),\qquad \beta=h(\tauS)+p_1\Sp(\tauS).
    \end{aligned}
    \label{eq:levels}
\end{equation}
Both levels grow linearly with the release count. Releasing more molecules therefore raises the desired signal and its own interference in the same proportion, and the ratio $r\triangleq\beta/\alpha=1+h(\tauS)/[p_1\Sp(\tauS)]$ is a property of the channel alone, which we call the \emph{ISI ratio}. It approaches one under severe interference, where the current pulse is small compared with the accumulated background, and grows large for a nearly ISI-free channel. Replacing the random interference history by its mean, as in \eqref{eq:levels}, is what makes a closed form possible.

\begin{remark}
\label{rem:tau}
The condition $\tauS<\Ts$ can fail, since for large $d$ or small $\Ds$ the pulse peak may arrive after the next release, leaving the decomposition \eqref{eq:rk} incomplete. The simulations include such conditions, since every release is simulated in full, and Section~\ref{sec:results} shows that every condition in which the channel fails for all release counts lies in this regime.
\end{remark}

\subsection{Ligand-Receptor Observation Model}
\label{sec:receptor}
The RX carries $\NR$ identical and independent receptors exposed to the same local ligand concentration, a description that covers biological and engineered receivers alike, since affinity-based biosensors observe the channel through the same finite-capacity binding process~\cite{Kuscu2016PhysicalDesign,Kuscu2021Graphene}. A receptor binds free ligand at rate $k_{\mathrm{on}}C$ and releases it at rate $k_{\mathrm{off}}$, and at equilibrium with a constant concentration $C$ it is occupied with the Langmuir probability
\begin{equation}
    q(C)=\frac{C}{\KD+C},\qquad \KD=\frac{k_{\mathrm{off}}}{k_{\mathrm{on}}},
    \label{eq:langmuir}
\end{equation}
which gives $q(\KD)=1/2$ and occupancy odds linear in concentration, $q/(1-q)=C/\KD$. Conditioned on the current bit, the sampled bound-receptor count is binomial,
\begin{equation}
    B_k\mid b_k=i\;\sim\;\Bin\!\big(\NR,q_i(\None)\big),\qquad q_i(\None)=q\big(C_i(\None)\big).
    \label{eq:binomial}
\end{equation}
Its variance $\NR q_i(1-q_i)$ is the binding noise, and the saturation is the vanishing sensitivity $\mathrm{d}q/\mathrm{d}C=\KD/(\KD+C)^2$ of the Langmuir function for $C\gg\KD$, which drives both occupancies toward one and collapses the mean count difference $\NR(q_1-q_0)$; for $C\ll\KD$ this difference, $\approx\NR(C_1-C_0)/\KD$, is instead masked by the binding noise.

Consecutive counts $B_{k-1}$ and $B_k$ are taken to be conditionally independent given their bits. This \emph{memoryless} assumption requires the receptor population to re-equilibrate between samples, which holds when the kinetic relaxation rate $k_{\mathrm{on}}C+k_{\mathrm{off}}$ is large compared with $1/\Ts$, as it does throughout Table~\ref{tab:params}, exceeding $1/\Ts$ by more than an order of magnitude at the base parameters. The assumption also neglects the correlation that consecutive samples inherit from their shared interference history. The simulations of Section~\ref{sec:methods} retain both effects.

\subsection{Comparator Detection}
\label{sec:comparator}
The receiver maintains no decision threshold and estimates no channel parameter. It decides from the direction of change of the bound-receptor count,
\begin{equation}
    \hat b_k=\begin{cases}1,&B_k>B_{k-1},\\[1pt] 0,&B_k<B_{k-1},\\[1pt] Z_k,&B_k=B_{k-1},\end{cases}
    \label{eq:comparator}
\end{equation}
where $Z_k$ is a fair coin flip independent of everything else. Rule \eqref{eq:comparator} is the receptor-count counterpart of the adaptive-threshold detector of~\cite{Damrath2016ATD} and is among the simplest detectors that a cell-scale receiver could implement~\cite{Kuscu2019ArchitecturesSurvey}.

\section{Optimal Release Count}
\label{sec:rule}

\subsection{Error Probability}
\label{sec:decomp}
Let $\Pe(\None)$ denote the bit error probability, or bit error rate (BER), of the comparator. Under the memoryless model of Section~\ref{sec:receptor}, the pair of bound-receptor counts $(B_{k-1},B_k)$ entering decision $k$ depends only on the ordered bit pair $(b_{k-1},b_k)$. In a transition $0\!\to\!1$ the counts $X_0\sim\Bin(\NR,q_0)$ and $X_1\sim\Bin(\NR,q_1)$, $q_1>q_0$, are independent, and the comparator fails when the count does not increase, which happens with probability
\begin{align}
    \Ptr(\None)&=\Prb\{X_1<X_0\}+\tfrac12\Prb\{X_1=X_0\}\nonumber\\
    &=\sum_{x_1=0}^{\NR}\phi_1(x_1)\Big[\sum_{x_0>x_1}\phi_0(x_0)+\tfrac12\phi_0(x_1)\Big],
    \label{eq:Ptr}
\end{align}
where $\phi_i(x)=\binom{\NR}{x}q_i^{x}(1-q_i)^{\NR-x}$. By exchangeability, the transition $1\!\to\!0$ fails equally often. For a repeated bit the two counts are i.i.d., and the fair-tie comparator fails with probability exactly $1/2$ whatever the release count.

\begin{proposition}[BER decomposition]
\label{prop:decomp}
For i.i.d.\ bits, the memoryless model \eqref{eq:binomial}, and the fair-tie comparator \eqref{eq:comparator},
\begin{equation}
    \Pe(\None)=\tfrac12\big(p_0^2+p_1^2\big)+2p_0p_1\,\Ptr(\None).
    \label{eq:decomp}
\end{equation}
Hence, for $0<p_1<1$, $\arg\min_{\None>0}\Pe=\arg\min_{\None>0}\Ptr$.
\end{proposition}
\begin{IEEEproof}
The ordered pairs $00,01,10,11$ occur with probabilities $p_0^2,p_0p_1,p_1p_0,p_1^2$ and are decided incorrectly with probabilities $1/2$, $\Ptr$, $\Ptr$, $1/2$, and the additive constant and the positive factor $2p_0p_1$ do not move the minimizer.
\end{IEEEproof}

The first term of \eqref{eq:decomp}, equal to $1/4$ for equiprobable bits, is the price of operating without an absolute decision threshold, since in the memoryless model the two counts of a repeated bit are identically distributed and a detector that only observes their change cannot do better than chance on them. This term is, however, not an error floor of the practical comparator, since in the actual channel the interference background drifts between consecutive samples, which makes the two counts of a repeated bit statistically different, and the error probabilities measured in Section~\ref{sec:results} accordingly drop well below $1/4$ under moderate and weak ISI. The location of the optimum, governed by Proposition~\ref{prop:decomp} through the transition term alone, is nevertheless captured faithfully by the model.

\subsection{Geometric-Mean Rule}
\label{sec:gm}
The transition error \eqref{eq:Ptr} decomposes per receptor. Each receptor contributes the difference of its current state $U\sim\mathrm{Bernoulli}(q_1)$ and its previous state $V\sim\mathrm{Bernoulli}(q_0)$, $\pm1$ if its state has changed and zero otherwise, and the comparator thus performs a majority decision over the state-changing receptors. Two properties of Langmuir binding make this majority decision analytically tractable.

\begin{lemma}[Constant sign bias]
\label{lem:bias}
For $C_0=\alpha\None$, $C_1=\beta\None$, $0<\alpha<\beta$, and independent $U\sim\mathrm{Bernoulli}(q_1)$, $V\sim\mathrm{Bernoulli}(q_0)$, the ratio $\Prb\{U-V=1\}/\Prb\{U-V=-1\}=q_1(1-q_0)/[q_0(1-q_1)]=C_1/C_0=\beta/\alpha$ does not depend on $\None$.
\end{lemma}
\begin{IEEEproof}
By \eqref{eq:langmuir}, $q_i/(1-q_i)=C_i/\KD$, and the ratio of the two odds, $C_1/C_0=\beta/\alpha$, contains neither $\KD$ nor $\None$.
\end{IEEEproof}

The fraction of state-changing receptors that change in the correct direction is therefore fixed by the channel. Define the \emph{activity} probability $\pA(\None)\triangleq\Prb\{U\neq V\}=q_1(1-q_0)+q_0(1-q_1)$ and the \emph{sign bias} $\rho\triangleq\Prb\{U-V=1\mid U\neq V\}=\beta/(\alpha+\beta)>1/2$. The count difference $X_1-X_0=\sum_{m=1}^{\NR}(U_m-V_m)$ has $K\sim\Bin(\NR,\pA)$ nonzero terms, each carrying the correct sign with the fixed probability $\rho$, and the release count therefore enters $\Ptr$ only through $\pA$.

\begin{lemma}[Monotonicity in the activity]
\label{lem:mono}
For fixed $\rho>1/2$ and $\NR\ge1$, the correct-decision probability $R_{\mathrm{tr}}(\pA)=1-\Ptr$ is strictly increasing in $\pA\in[0,1)$.
\end{lemma}
\begin{IEEEproof}
Given $K=k$ active receptors, let $Y_k\sim\Bin(k,\rho)$ count those carrying the correct sign. With fair tie-breaking the comparator decides correctly with probability $G(k)=\Prb\{Y_k>k/2\}+\tfrac12\Prb\{Y_k=k/2\}$, $G(0)=\tfrac12$. Conditioning on one additional active receptor gives $G(2j+1)-G(2j)=(\rho-\tfrac12)\Prb\{Y_{2j}=j\}>0$ for $j\ge0$ and $G(2j)-G(2j-1)=0$ for $j\ge1$, the latter by the identity $(1-\rho)\Prb\{Y_{2j-1}=j\}=\rho\Prb\{Y_{2j-1}=j-1\}$, hence $G$ is nondecreasing with strictly positive increments at odd $k$. With $K\sim\Bin(\NR,\pA)$, $R_{\mathrm{tr}}(\pA)=\E[G(K)]$ is the Bernstein polynomial of $G$, whose derivative $\NR\sum_{k=0}^{\NR-1}\binom{\NR-1}{k}\pA^k(1-\pA)^{\NR-1-k}[G(k+1)-G(k)]$ has only nonnegative terms, the $k=0$ term equaling $\NR(1-\pA)^{\NR-1}(\rho-\tfrac12)>0$ for $\pA<1$.
\end{IEEEproof}

\begin{lemma}[Unique maximizer of the activity]
\label{lem:max}
Under \eqref{eq:levels} and \eqref{eq:langmuir}, $\pA(\None)$ is uniquely maximized over $\None>0$ at $\KD/\sqrt{\alpha\beta}$.
\end{lemma}
\begin{IEEEproof}
Substituting \eqref{eq:langmuir} into the activity gives
\begin{equation}
    \pA(\None)=\frac{\KD(\alpha+\beta)\None}{(\KD+\alpha\None)(\KD+\beta\None)},
    \label{eq:pA_N}
\end{equation}
whose logarithmic derivative $(\KD^2-\alpha\beta\None^2)/[\None(\KD+\alpha\None)(\KD+\beta\None)]$ is positive below $\KD/\sqrt{\alpha\beta}$, zero there, and negative beyond.
\end{IEEEproof}

\begin{theorem}[Comparator-optimal release count]
\label{thm:main}
Under the two-level memoryless binomial model with Langmuir binding, i.i.d.\ bits with $0<p_1<1$, and the fair-tie comparator, the bit error probability is uniquely minimized over $\None>0$ at
\begin{equation}
    \boxed{\;\Nstar=\frac{\KD}{\sqrt{\alpha\beta}}\;}\qquad\Longleftrightarrow\qquad \KD=\sqrt{C_0(\Nstar)\,C_1(\Nstar)}.
    \label{eq:Nstar}
\end{equation}
\end{theorem}
\begin{IEEEproof}
By Proposition~\ref{prop:decomp} it suffices to minimize $\Ptr$, which by Lemma~\ref{lem:mono} strictly decreases in $\pA$, and by Lemma~\ref{lem:max} $\pA$ has the unique maximizer $\KD/\sqrt{\alpha\beta}$, at which $\KD=\sqrt{C_0C_1}$ since $C_0C_1=\alpha\beta\None^2$.
\end{IEEEproof}

\begin{figure}[!t]
\centering
\includegraphics[width=\columnwidth]{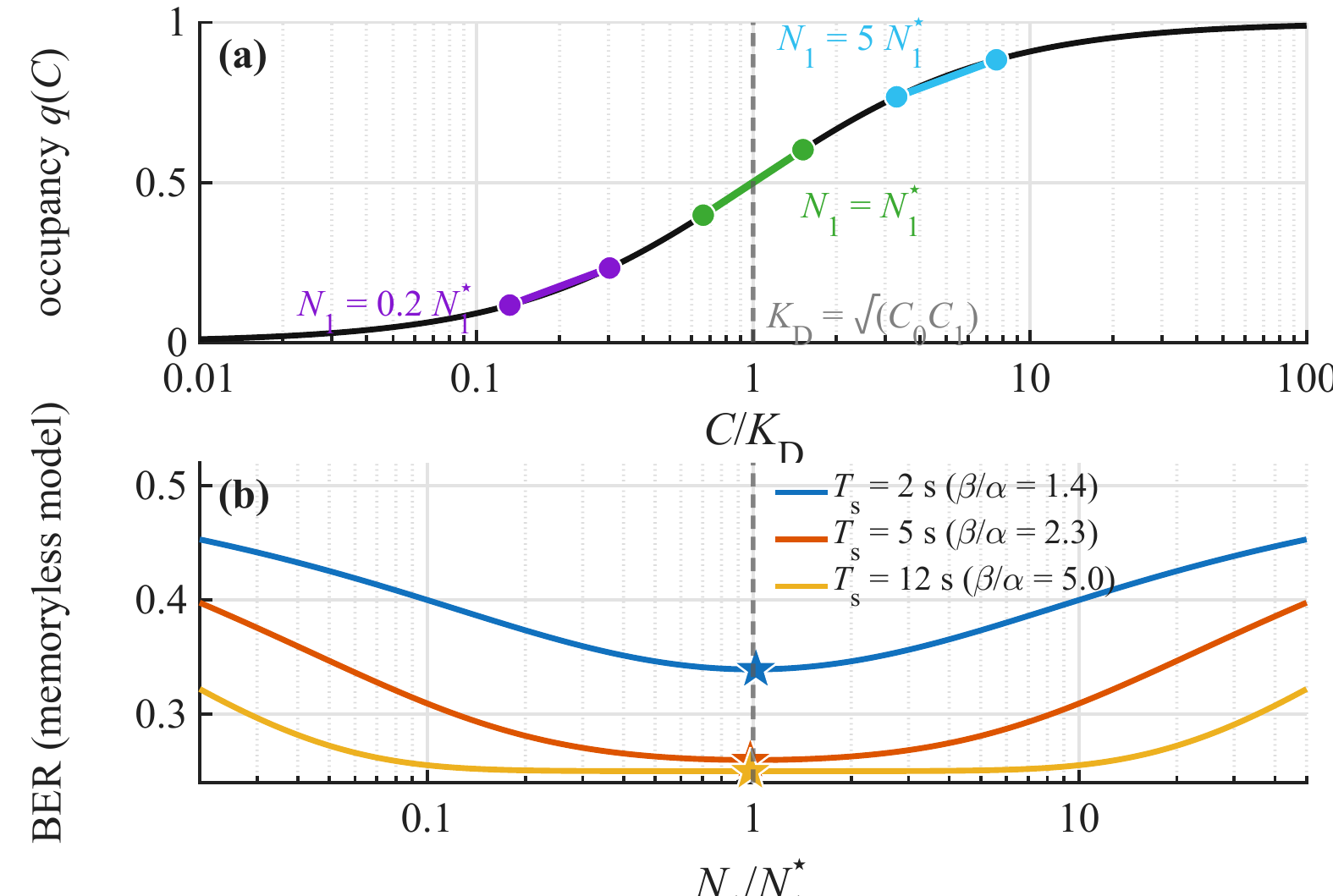}
\caption{The geometric-mean rule. (a) Langmuir occupancy with the bit-conditioned levels $C_0=\alpha\None$ and $C_1=\beta\None$ ($\beta/\alpha=2.3$) for release counts below, at, and above $\Nstar$, where the levels straddle $\KD$ log-symmetrically on the steepest part of the curve. (b) Exact BER of the memoryless model versus the normalized release count for three ISI ratios ($\NR=50$, $p_1=0.5$), stars marking the minima.}
\label{fig:concept}
\end{figure}

Fig.~\ref{fig:concept}(a) illustrates the optimality condition. On a logarithmic concentration axis, the optimal transmitter places the two levels symmetrically about the dissociation constant, on the steepest part of the Langmuir curve, the same operating condition that our previous work~\cite{Kuscu2026AdaptiveReceiver} reached from the receiver side by retuning the receptor to given levels. Fig.~\ref{fig:concept}(b) shows the exact model error probability from \eqref{eq:Ptr} and \eqref{eq:decomp}, a U-shaped curve in the normalized release count with its minimum at $\Nstar$ for every ISI condition.

\subsection{Universality and Robustness}
\label{sec:properties}
Write $\None=\mu\Nstar$. Then $C_0/\KD=\mu/\sqrt r$ and $C_1/\KD=\mu\sqrt r$ with $r=\beta/\alpha$, and since the binomial parameters \eqref{eq:binomial} depend only on $C_i/\KD$, the following holds.

\begin{corollary}[Universality]
\label{cor:universal}
As a function of $\mu=\None/\Nstar$, the model bit error probability depends on the channel only through the ISI ratio $r$, besides $\NR$ and $p_1$.
\end{corollary}

Distance, diffusivity, and symbol duration therefore act on the normalized error curve only through the single scalar $r$, the receptor number deepens the minimum without moving it, and $\KD$ enters only through the linear scaling $\Nstar\propto\KD$. Moreover, $\pA$ in \eqref{eq:pA_N} is invariant under $\mu\to1/\mu$, so the model error curve is symmetric about $\Nstar$ on a logarithmic axis (Fig.~\ref{fig:concept}(b)), and releasing a factor $\kappa$ too many molecules costs exactly as much as releasing a factor $\kappa$ too few.

From \eqref{eq:Nstar}, $\partial\ln\Nstar/\partial\ln\KD=1$ and $\partial\ln\Nstar/\partial\ln\alpha=\partial\ln\Nstar/\partial\ln\beta=-1/2$, hence a twofold error in either level coefficient displaces the operating point by only a factor of $\sqrt2$, well within the flat region around the minimum of the error curves observed in Section~\ref{sec:results}. Since $\Nstar$ is in general not an integer, the optimal integer count is $\lfloor\Nstar\rfloor$ or $\lceil\Nstar\rceil$, the latter exactly when $\KD^2\ge\alpha\beta\lfloor\Nstar\rfloor(\lfloor\Nstar\rfloor+1)$ by \eqref{eq:pA_N}, a distinction that is immaterial at the release counts of practical interest.

\begin{remark}[Direction of the residual offset]
\label{rem:history}
The memoryless model averages the random interference history. A first refinement conditions on the previous bit as well. With $g_1=h(\tauS+\Ts)$ the contribution of the immediately preceding symbol to the current sample, the mean levels seen by the previous and the current sample depend on the ordered bit pair and equal $(\alpha,\,\alpha-p_1g_1)\None$, $(\alpha,\,\beta-p_1g_1)\None$, $(\beta,\,\alpha+(1-p_1)g_1)\None$, and $(\beta,\,\beta+(1-p_1)g_1)\None$ for the pairs $00$, $01$, $10$, and $11$. Each pair has a level ratio independent of $\None$, and Lemmas~\ref{lem:bias} to~\ref{lem:max} therefore apply to each pair separately, yielding a closed-form optimum $\KD/\sqrt{\ell_{\mathrm{prev}}\ell_{\mathrm{cur}}}$ per pair in its two level coefficients $\ell_{\mathrm{prev}}$ and $\ell_{\mathrm{cur}}$. These four values bracket the refined optimum, and $\Nstar$ lies inside the bracket. As ISI weakens, the repeated-zero pair, whose optimum $\KD/\sqrt{\alpha(\alpha-p_1g_1)}$ exceeds $\Nstar$, dominates the error probability, and the refined optimum moves slightly above $\Nstar$ but never below it. This explains the direction of the small offsets between predicted and empirically optimal release counts in Section~\ref{sec:results}, and shows that $\Nstar$ is, if anything, slightly conservative.
\end{remark}

\subsection{Closed-Form Evaluation of the ISI Coefficient}
\label{sec:closedform}
The slopes $\alpha$ and $\beta$ involve the infinite interference sum $\Sp(\tauS)$ of \eqref{eq:spast}, which converges too slowly for term-by-term evaluation on a resource-limited device, since truncation after even $2\times10^4$ terms still underestimates the sum by about one percent. The change of variable $z=d/(2\sqrt{\Ds t})$ yields the exact tail identity $\int_{t_0}^{\infty}h\dd t=\erfop\big(d/(2\sqrt{\Ds t_0})\big)/(4\pi\Ds d)$ for $t_0>0$, and the Euler--Maclaurin formula, $\sum_{\ell=0}^{\infty}f(t_0+\ell\Ts)=\Ts^{-1}\int_{t_0}^{\infty}f\dd t+\tfrac12f(t_0)-\tfrac{\Ts}{12}f'(t_0)+O\big(\Ts^3 f'''(t_0)\big)$, converts the discrete tail of \eqref{eq:spast} into this integral plus local corrections. Retaining the first $L$ terms of \eqref{eq:spast} exactly and applying the formula to the remainder from $t_L=\tauS+(L+1)\Ts$ gives, with $h'(t)=h(t)\big(-\tfrac{3}{2t}+\tfrac{d^2}{4\Ds t^2}\big)$,
\begin{equation}
    \begin{aligned}
    \hat S^{(L)}(\tauS)={}&\sum_{\ell=1}^{L}h(\tauS+\ell\Ts)+\frac1\Ts\int_{t_L}^{\infty}h(t)\dd t\\
    &+\frac12h(t_L)-\frac{\Ts}{12}h'(t_L).
    \end{aligned}
    \label{eq:Shyb}
\end{equation}
Beyond its peak $h$ is smooth on the scale of $\Ts$, and against a reference obtained by summing $2\times10^6$ terms exactly plus the Euler--Maclaurin remainder far in the tail, \eqref{eq:Shyb} is accurate over $\Ts/t_{\mathrm{peak}}\in[1,30]$ and the geometries of Table~\ref{tab:params} to within $4\times10^{-3}$ relative error at $L=0$, $5\times10^{-5}$ at $L=2$, and below $10^{-5}$ at the working value $L=5$, whereas the bare integral approximation, the first term of \eqref{eq:S0} below, deviates by up to $22\%$. Substituting \eqref{eq:Shyb} into \eqref{eq:levels} and \eqref{eq:Nstar} yields the release count as an explicit function of the channel parameters,
\begin{equation}
    \boxed{\;\Nstar=\frac{\KD}{\sqrt{p_1\hat S^{(L)}(\tauS)\big[h(\tauS)+p_1\hat S^{(L)}(\tauS)\big]}}\;,}
    \label{eq:Nstar_closed}
\end{equation}
which is computable from the six quantities $\KD$, $p_1$, $\tauS$, $\Ts$, $d$, and $\Ds$ with a handful of exponentials and one error function.

A fully elementary variant follows by setting $L=0$ in \eqref{eq:Shyb}, which removes the summation and leaves, with $t_0=\tauS+\Ts$,
\begin{equation}
    \hat S^{(0)}(\tauS)=\frac{\erfop\!\big(\tfrac{d}{2\sqrt{\Ds t_0}}\big)}{4\pi\Ds d\,\Ts}
    +\frac{h(t_0)}{2}-\frac{\Ts\,h'(t_0)}{12}.
    \label{eq:S0}
\end{equation}
This variant requires a single evaluation of $h$ and one error function and is therefore suited to the most resource-limited transmitters. Its relative error stays below $0.4\%$ over the studied range, which through the square root in \eqref{eq:Nstar_closed} changes $\Nstar$ by less than $0.2\%$, far below the resolution of the simulations reported below, and the two forms can therefore be used interchangeably by resource-limited TXs in practice.

\section{Simulation Framework}
\label{sec:methods}

We evaluate \eqref{eq:Nstar_closed} against the physical effects excluded by its assumptions using two simulation layers of increasing fidelity. A time-domain Monte Carlo (TD-MC) simulator generates the continuous-time channel waveforms, restoring the random interference history, in which every symbol experiences the ISI of its actual predecessors, and the finite-rate binding kinetics of the receptors. The particle-based reaction-diffusion simulator Smoldyn~\cite{Andrews2017Smoldyn} additionally represents discrete molecules, the finite spherical-receiver geometry, and spatially resolved stochastic binding. Neither layer uses the two-level abstraction of Section~\ref{sec:sampling}. Table~\ref{tab:params} lists the parameters, chosen for a cell-scale biosensing setting. One molecule per $\um^3$ corresponds to $1.66$ nM, and the swept dissociation constants $\KD=0.1$ to $5\,\um^{-3}$ span roughly $0.2$ to $8$ nM, the affinity class of antibody- and aptamer-based biosensing receivers~\cite{Kuscu2016PhysicalDesign,Kuscu2021Graphene}. The binding rate $k_{\mathrm{on}}=10\,\um^3$/s, about $6\times10^{9}$ M$^{-1}$s$^{-1}$, represents fast, diffusion-limited association~\cite{Berg1977Chemoreception}, and the resulting unbinding rates of $1$ to $50$ s$^{-1}$ keep receptor relaxation faster than the symbol durations. The diffusion coefficients cover macromolecular ligands in aqueous and crowded biological media, and the distances of $10$ to $40\,\um$ correspond to one to a few cell lengths.

\begin{table}[!t]
\centering
\caption{Simulation parameters (default values; swept ranges in parentheses)}
\label{tab:params}
\scriptsize
\begin{tabular}{@{}l c p{3.3cm}@{}}
\toprule
Parameter & Symbol & Value \\
\midrule
TX-RX distance; RX radius & $d$; $r_{\mathrm{rx}}$ & $20\,\um$ ($10$--$40\,\um$); $2\,\um$ \\
Ligand diffusion coefficient & $\Ds$ & $10\,\um^2$/s ($2$--$40$) \\
Symbol period & $\Ts$ & $5$ s ($2$--$40$ s) \\
Sampling phase; bit prior & $\tauS$; $p_1$ & $t_{\mathrm{peak}}=d^2/(6\Ds)$; $0.5$ \\
Number of receptors & $\NR$ & $50$ ($10$--$150$) \\
Dissociation constant & $\KD$ & $0.5\,\um^{-3}$ ($0.1$--$5$) \\
Binding / unbinding rates & $k_{\mathrm{on}}$, $k_{\mathrm{off}}$ & $10\,\um^3$/s, $\KD k_{\mathrm{on}}$ \\
Tested release counts & $\None/\Nstar$ & $0.01$, $0.05$, $0.1$, $0.5$, $1$, $1.5$, $2$, $5$, $10$ \\
Time step; run length & $\Delta t$; $T_{\mathrm{tot}}$ & $0.01$ s; $2000$ s \\
Runs per point& --- & $100$ (TD-MC), $15$ (Smoldyn) \\
\bottomrule
\end{tabular}
\end{table}

\subsection{Time-Domain Monte Carlo Simulations}
\label{sec:tdmc}
Each TD-MC run draws a fresh random bit sequence of $\lfloor T_{\mathrm{tot}}/\Ts\rfloor$ symbols and superposes the Green's-function responses \eqref{eq:superposition} of all releases on a time grid of step $\Delta t$. The receptor occupancy is propagated through the binding kinetics $\dot p_{\mathrm{B}}=k_{\mathrm{on}}\,r(t)\,(1-p_{\mathrm{B}})-k_{\mathrm{off}}\,p_{\mathrm{B}}$ on the same grid, which retains the finite response time that the equilibrium model neglects. At each sampling instant $k\Ts+\tauS$, a count $B_k\sim\Bin(\NR,p_{\mathrm{B}})$ is drawn and the comparator \eqref{eq:comparator} is applied. The first quarter of every run is discarded to let the ISI build up, and every tested release count is evaluated over $100$ independent runs. We report the mean BER and its standard error of the mean (SEM) across runs.

\subsection{Particle-Based Simulations}
\label{sec:smoldyn}
\begin{figure}[!t]
\centering
\includegraphics[width=\columnwidth]{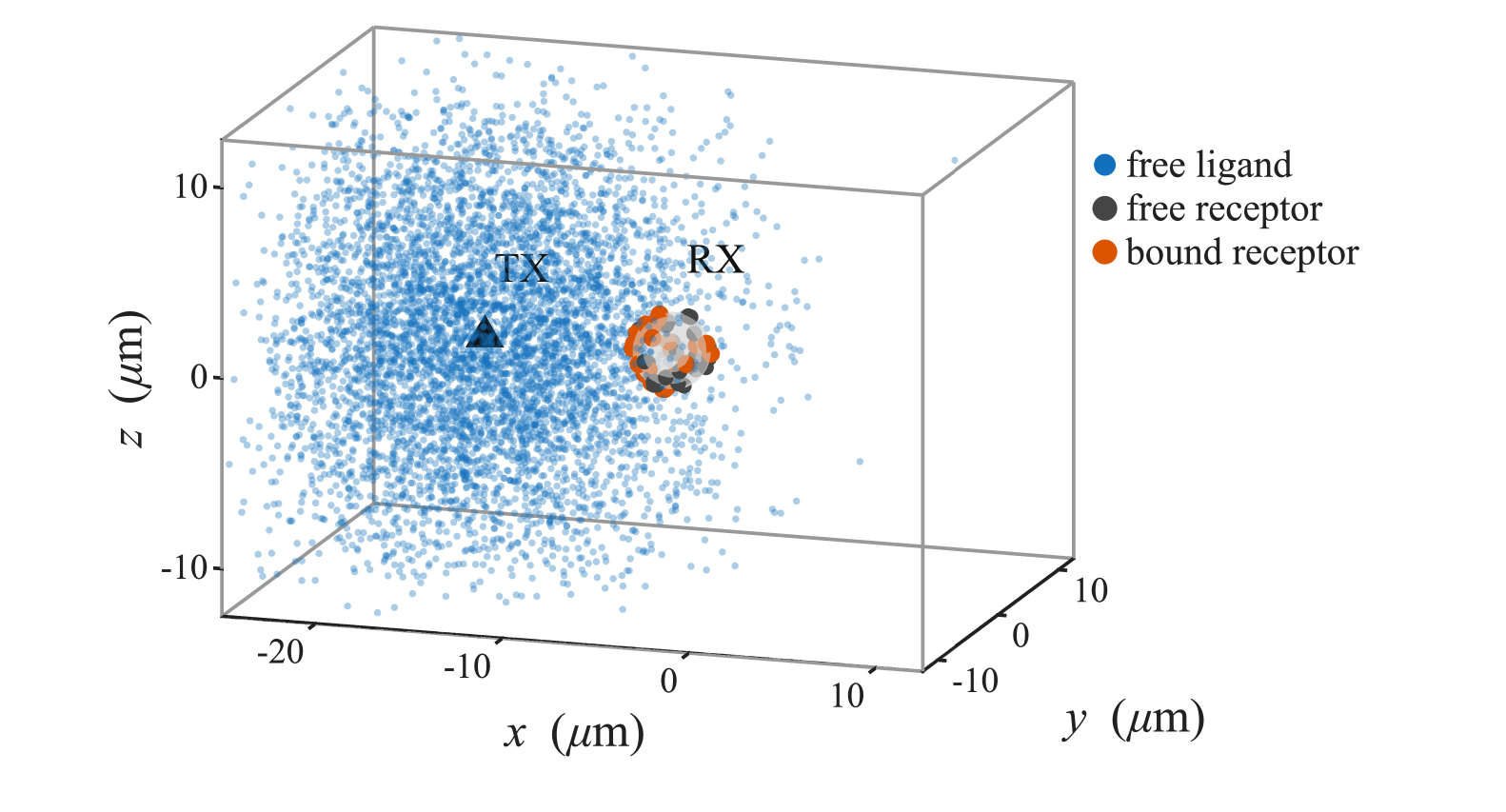}
\caption{The Smoldyn arena, from the recorded molecule positions of a $d=10\,\um$ run about $t_{\mathrm{peak}}$ after a bit-1 release of $\Nstar$ molecules. Blue dots are free ligands diffusing from the TX toward the receiver sphere, which carries free (gray) and ligand-bound (orange) receptors.}
\label{fig:arena}
\end{figure}

In Smoldyn, we represent the receiver as a reflecting sphere of radius $r_{\mathrm{rx}}$ carrying $\NR$ surface receptors at distance $d$ from the release point (TX), inside a box with unbounded-emitter boundaries (Fig.~\ref{fig:arena}). The TX releases $\None$ ligands for each bit 1, which diffuse with coefficient $\Ds$ and bind free surface receptors at rate $k_{\mathrm{on}}$, unbinding at rate $k_{\mathrm{off}}$. The number of complexes is sampled at $k\Ts+\tauS$, after which detection and BER evaluation proceed exactly as in TD-MC, and Smoldyn run $r$ replays the bit sequence of TD-MC run $r$ in a paired design. Particle-based simulation is expensive, since the cost of each of the $2\times10^5$ time steps of a run scales with the number of molecules in flight, up to hundreds of thousands here, whereas a TD-MC run propagates one mean-field kinetic equation on the same grid. Smoldyn was therefore applied to thirteen selected conditions spread over all four sweeps (Table~\ref{tab:smoldyn}), with $15$ runs per release count. 

\subsection{Cross-Validation of the Simulation Layers}
\label{sec:trace}
\begin{figure*}[!t]
\centering
\includegraphics[width=\textwidth]{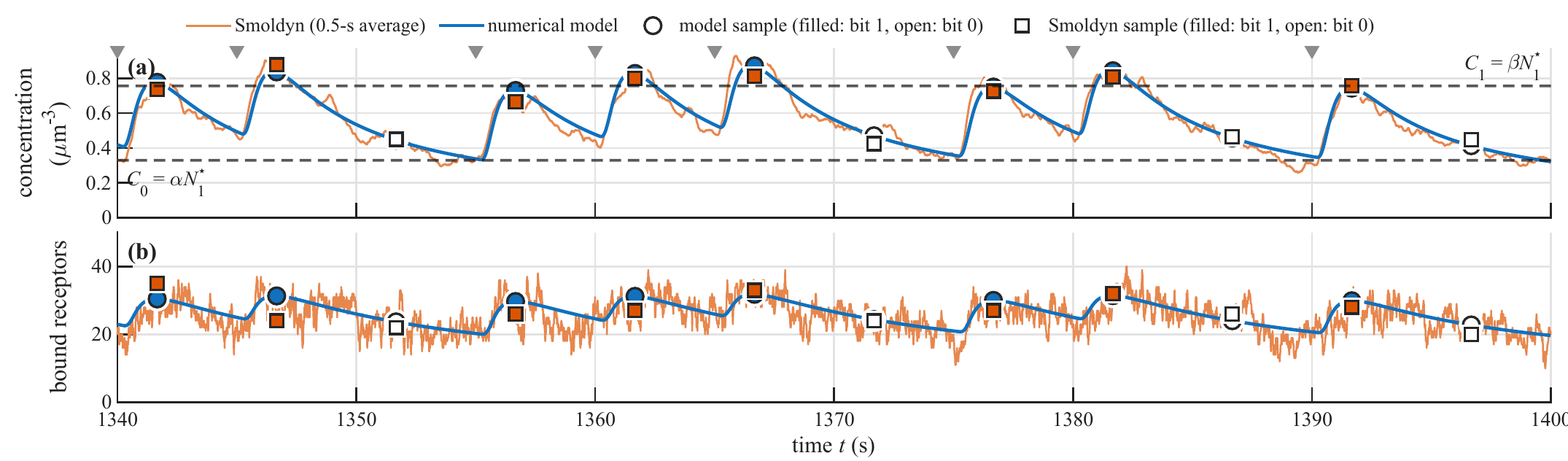}
\caption{Cross-validation on raw time traces ($d=10\,\um$, $\Ts=5$ s, $\None=\Nstar$, other parameters at base values; gray triangles at the top edge mark the bit-1 release instants). (a) Ligand concentration near the receiver, Smoldyn (orange) versus the Green's-function superposition driving TD-MC (blue), with the asymptotic bit-conditioned levels via \eqref{eq:Shyb} dashed. Circles and squares mark the model and Smoldyn samples at the detection instants, filled for bit-1 and open for bit-0. (b) Bound-receptor count, Smoldyn (orange) versus the kinetic mean-field trajectory $\NR\,p_{\mathrm{B}}(t)$ (blue), samples marked likewise.}
\label{fig:trace}
\end{figure*}

Fig.~\ref{fig:trace} verifies that the numerical channel model driving TD-MC, the particle-based physics, and the two-level abstraction of Section~\ref{sec:sampling} are mutually consistent, showing a $60$-s segment from a Smoldyn run in the $d=10\,\um$ configuration with $\Ts=5$ s at $\None=\Nstar$, with the Green's-function reconstruction of the same bit sequence. The measured ligand concentration tracks the superposition pulse by pulse with the counting noise of discrete molecules, the normalized root-mean-square (RMS) deviation of the smoothed trace staying below seven percent, and the bound-receptor count fluctuates around the kinetic mean-field trajectory $\NR\,p_{\mathrm{B}}(t)$ with a standard deviation consistent with the binomial value $\sqrt{\NR\,p_{\mathrm{B}}(1-p_{\mathrm{B}})}$ of independent receptors. Above all, the concentrations sampled during bit-0 and bit-1 symbols cluster, in both layers, around the asymptotic levels $C_0=\alpha\Nstar$ and $C_1=\beta\Nstar$ of \eqref{eq:levels}, drawn as dashed lines. The two-level abstraction with the closed-form ISI coefficient \eqref{eq:Shyb} thus describes the channel quantitatively.

\begin{figure*}[!t]
\centering
\includegraphics[width=\textwidth]{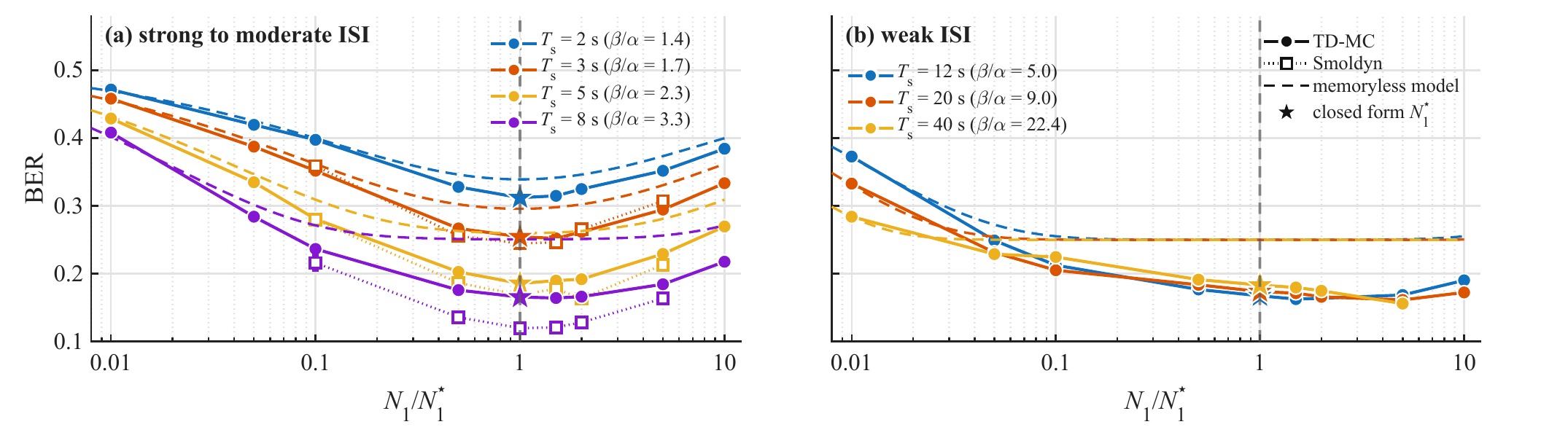}
\caption{BER versus the normalized release count for the $d=10\,\um$ family ($\tauS=t_{\mathrm{peak}}=1.67$ s, other parameters at base values), under (a) strong to moderate and (b) weak ISI; TD-MC is the mean $\pm$ SEM of $100$ runs, Smoldyn of $15$ runs.}
\label{fig:base}
\end{figure*}

\subsection{Simulation Study and Metrics}
\label{sec:metrics}
The study pairs the symbol period with one physical parameter, others at default values, in four two-dimensional sweeps of the receptor affinity ($\KD\in\{0.1,0.2,0.5,1,5\}\,\um^{-3}$), the receptor number ($\NR\in\{10,20,30,50,75,100,150\}$), the diffusion coefficient ($\Ds\in\{2,5,10,20,40\}\,\um^2$/s), and the distance ($d\in\{10,15,20,25,30,40\}\,\um$), each against $\Ts\in\{2,3,5,8,12,20\}$ s, extended to $40$ s for affinity and distance. Since the default configuration belongs to all four sweeps, their $149$ entries comprise $130$ distinct conditions. Fig.~\ref{fig:matrix} reports all entries and Figs.~\ref{fig:base} and~\ref{fig:sweeps} representative slices. In every condition the closed form \eqref{eq:Nstar_closed} is evaluated first, the BER is then measured over the common grid of release counts in Table~\ref{tab:params}, expressed as multiples of $\Nstar$, and the prediction is marked by a star ($\star$) throughout, so that the rule holds wherever the star sits at or beside the minimum.

Each BER curve is summarized by the location $N_1^{\min}/\Nstar$ of its empirical minimum and by the penalty $\gamma\triangleq\Pe(\Nstar)/\Pe^{\min}-1$, the relative increase in error probability from operating at the closed-form count instead of the empirically best one, where $\Pe^{\min}$ is the minimum BER over the tested grid. Conditions in which even the best release count leaves the BER above $0.45$ are excluded from the penalty statistics, since the channel then conveys essentially no information at any $\None$ and an optimal release count loses its meaning. Isolated points at the largest release multiples, where the binding rate $k_{\mathrm{on}}C_1$ drives the explicit kinetics integrator out of its stability region, are excluded.

\section{Results}
\label{sec:results}

\subsection{Symbol-Period Dependence}
\label{sec:res_base}
Fig.~\ref{fig:base} presents the main result for the $d=10\,\um$ family, where $\tauS=t_{\mathrm{peak}}=1.67$ s satisfies $\tauS<\Ts$ at every symbol period and Smoldyn coverage is densest. The symbol period runs from strong ISI ($\Ts=2$ s, $\beta/\alpha=1.4$) to nearly none ($\Ts=40$ s, $\beta/\alpha=22$). First, every curve is U-shaped. Moving from $0.01\Nstar$ toward the star, the BER falls steeply, by up to a factor of three, as the growing release count lifts the two receptor occupancies out of the binding-noise floor, and past the optimum it climbs again as saturation sets in. The U-shape spans the three decades of the tested grid, and at $10\Nstar$ the BER is again comparable to that at $0.1\Nstar$, so a transmitter that simply maximizes its release count performs as poorly as one releasing ten times too few.

\begin{figure*}[!t]
\centering
\includegraphics[width=\textwidth]{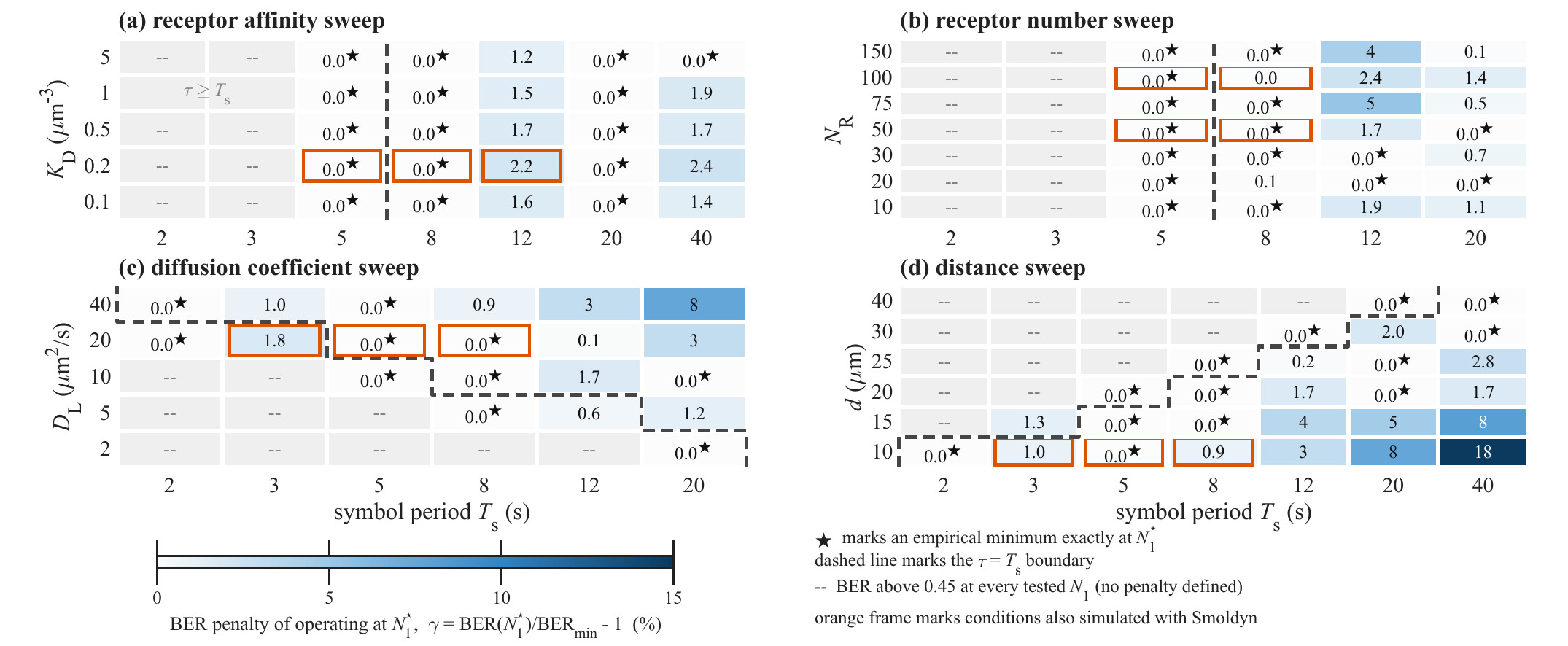}
\caption{The validation matrix over all four sweeps, the default-configuration row recurring in every panel. Each cell reports the BER penalty $\gamma$ (in percent) of operating at the closed-form prediction instead of the empirically best tested release count, with the markings explained in the key. Gray cells, in which the BER exceeds $0.45$ at every tested release count, lie entirely beyond the model-validity boundary $\tauS=\Ts$. The thirteen orange-framed conditions are those of Table~\ref{tab:smoldyn}.}
\label{fig:matrix}
\end{figure*}

Second, the predictions sit at the minima. For all symbol periods with strong to moderate ISI ($\Ts\le8$ s), the best tested release count is $\Nstar$ or the adjacent $1.5\Nstar$, and the BER at the predicted count lies within one percent of the minimum (Fig.~\ref{fig:matrix}(d)). The Smoldyn results (open squares) reproduce the minimum's location at $1$ to $2\,\Nstar$ with a slightly lower BER, since the finite particle domain cannot sustain the full diffusive tail of the unbounded medium, and the runs carry slightly less ISI, which by Remark~\ref{rem:history} nudges the empirical optimum marginally upward, never below the prediction.

Third, the memoryless model (dashed) predicts the minimum's location even when its BER estimates differ from simulation. Under weak ISI its repeated-bit term keeps the model BER near $1/4$ while the physical comparator, aided by the interference drift between samples, performs better. Only under weak ISI ($\Ts\ge12$ s) does the minimum drift upward, to between $2\Nstar$ and $5\Nstar$, as Remark~\ref{rem:history} predicts, and the curve is then flat enough that using $\Nstar$ costs little in absolute terms, a BER increase of at most $0.03$ at $\Ts=40$ s.

\begin{figure*}[!t]
\centering
\includegraphics[width=\textwidth]{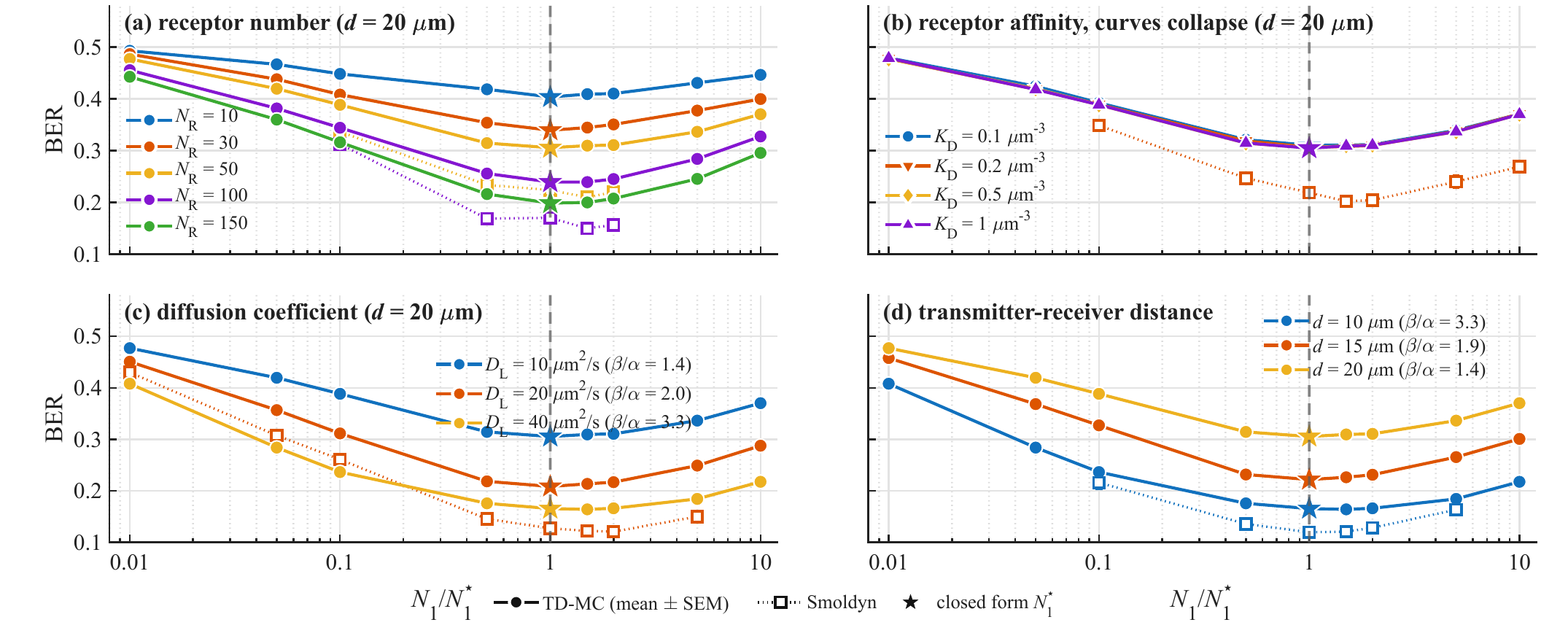}
\caption{BER versus the normalized release count at $\Ts=8$ s under sweeps of (a) the receptor number, (b) the receptor affinity, (c) the diffusion coefficient, and (d) the transmitter-receiver distance, other parameters at base values. In (b) the four curves collapse, confirming the predicted scale invariance in $\KD$.}
\label{fig:sweeps}
\end{figure*}

\subsection{Parameter Sweeps}
\label{sec:res_sweeps}
Fig.~\ref{fig:sweeps} repeats the comparison at $\Ts=8$ s while sweeping, in turn, the receptor number, the receptor affinity, the diffusion coefficient, and the distance, tracing out the universality structure of Corollary~\ref{cor:universal}. Increasing the receptor number from $10$ to $150$ (Fig.~\ref{fig:sweeps}(a)) halves the BER at the optimum but leaves the minimum at the predicted release count. Sweeping the affinity across an order of magnitude (Fig.~\ref{fig:sweeps}(b)) is a stronger test, since $\Nstar$ scales linearly with $\KD$ while the normalized curve does not depend on $\KD$ at all. The four TD-MC curves collapse onto one another within Monte Carlo error, confirming the predicted scale invariance, and the Smoldyn points at $\KD=0.2\,\um^{-3}$ follow the same collapsed curve. Sweeping the diffusion coefficient (Fig.~\ref{fig:sweeps}(c)) and the distance (Fig.~\ref{fig:sweeps}(d)) changes the ISI ratio and with it both the depth of the minimum and the absolute value of $\Nstar$, the latter by more than an order of magnitude across the diffusion-coefficient sweep, yet in every case the prediction remains at, or one grid point beside, the empirical minimum.

\subsection{BER Penalty Across All Conditions}
\label{sec:res_matrix}
Fig.~\ref{fig:matrix} reports all $149$ sweep entries, each cell annotated with the penalty $\gamma$, and the statistics below count each of the $130$ distinct conditions once. Of these, $87$ reach a BER below $0.45$ at their best release count and enter the penalty statistics. Across them, the median penalty is $0.0\%$, more than $90\%$ are below $3.5\%$, and $46$ minima occur at $\Nstar$ and $79$ within $[\Nstar,2\Nstar]$. No minimum lies below the prediction in either simulation method, consistent with the one-sided offset of Remark~\ref{rem:history}. In the ISI-relevant regime $\beta/\alpha\le4$ the agreement is tighter still, all $75$ such conditions having their minimum within $[\Nstar,2\Nstar]$ and a worst-case penalty of $5.3\%$. The few larger penalties, up to $17.5\%$, all belong to the nearly ISI-free regime $\beta/\alpha>4$, where the BER curve is flat around its minimum and even a suboptimal release count costs little in absolute terms. The gray cells, with BER near $0.5$ for every tested release count over three orders of magnitude, all lie beyond the $\tauS=\Ts$ boundary, whereas several conditions just beyond it remain usable, thus $\tauS<\Ts$ is a sufficient but conservative feasibility check.

Table~\ref{tab:smoldyn} details the thirteen conditions verified by particle-based simulation, which span all four sweeps, ISI ratios from $1.3$ to $3.3$, and predicted counts $\Nstar$ from $3.7\times10^{3}$ to $3.8\times10^{4}$ molecules. The empirical minimum lies at $1$ to $2\,\Nstar$ in every condition, and the penalty for operating at the prediction ranges from zero to about fourteen percent with a median below five percent, values that include the scatter of the fifteen-run ensembles. Because they reproduce both the location of the optimum and the shape of the BER curve around it, the particle simulations corroborate the TD-MC results in the remaining conditions, where particle simulation would be prohibitive.

\begin{table}[!t]
\centering
\caption{Particle-Based (Smoldyn) Validation Conditions}
\label{tab:smoldyn}
\scriptsize
\setlength{\tabcolsep}{2pt}
\begin{tabular}{@{}lcccc@{\hspace{4pt}}lcccc@{}}
\toprule
Condition & $\Ts$ & $\beta/\alpha$ & $\gamma$ & $\frac{N_1^{\min}}{\Nstar}$ & Condition & $\Ts$ & $\beta/\alpha$ & $\gamma$ & $\frac{N_1^{\min}}{\Nstar}$ \\
\midrule
$d$=10 $\mu$m & 3 & 1.71 & 0.0\% & 1 & $\NR$=50 & 8 & 1.45 & 6.8\% & 1.5 \\
$d$=10 $\mu$m & 5 & 2.29 & 3.3\% & 2 & $\NR$=100\textsuperscript{$\dagger$} & 5 & 1.27 & 5.4\% & 2 \\
$d$=10 $\mu$m & 8 & 3.33 & 0.0\% & 1 & $\NR$=100 & 8 & 1.45 & 13.6\% & 1.5 \\
$\Ds$=20 $\mu$m$^2$/s\textsuperscript{$\dagger$} & 3 & 1.33 & 1.2\% & 1.5 & $\KD$=0.2 $\mu$m$^{-3}$\textsuperscript{$\dagger$} & 5 & 1.27 & 3.2\% & 1.5 \\
$\Ds$=20 $\mu$m$^2$/s & 5 & 1.57 & 4.4\% & 1.5 & $\KD$=0.2 $\mu$m$^{-3}$ & 8 & 1.45 & 8.3\% & 1.5 \\
$\Ds$=20 $\mu$m$^2$/s & 8 & 1.99 & 5.5\% & 2 & $\KD$=0.2 $\mu$m$^{-3}$ & 12 & 1.71 & 4.7\% & 1.5 \\
$\NR$=50\textsuperscript{$\dagger$} & 5 & 1.27 & 5.1\% & 1.5 & & & & & \\
\bottomrule
\multicolumn{10}{@{}p{8.6cm}@{}}{\rule{0pt}{2ex}\scriptsize Other parameters at base values; $\Ts$ in s. $\beta/\alpha$ is the ISI ratio, $\gamma=\Pe(\Nstar)/\Pe^{\min}-1$ the penalty of operating at $\Nstar$, and $N_1^{\min}$ the location of the empirical minimum. \textsuperscript{$\dagger$}$\tau$ slightly larger than $\Ts$.}
\end{tabular}
\end{table}

\subsection{Effect of Suboptimal Release Counts}
\label{sec:res_gains}
\begin{figure}[!t]
\centering
\includegraphics[width=\columnwidth]{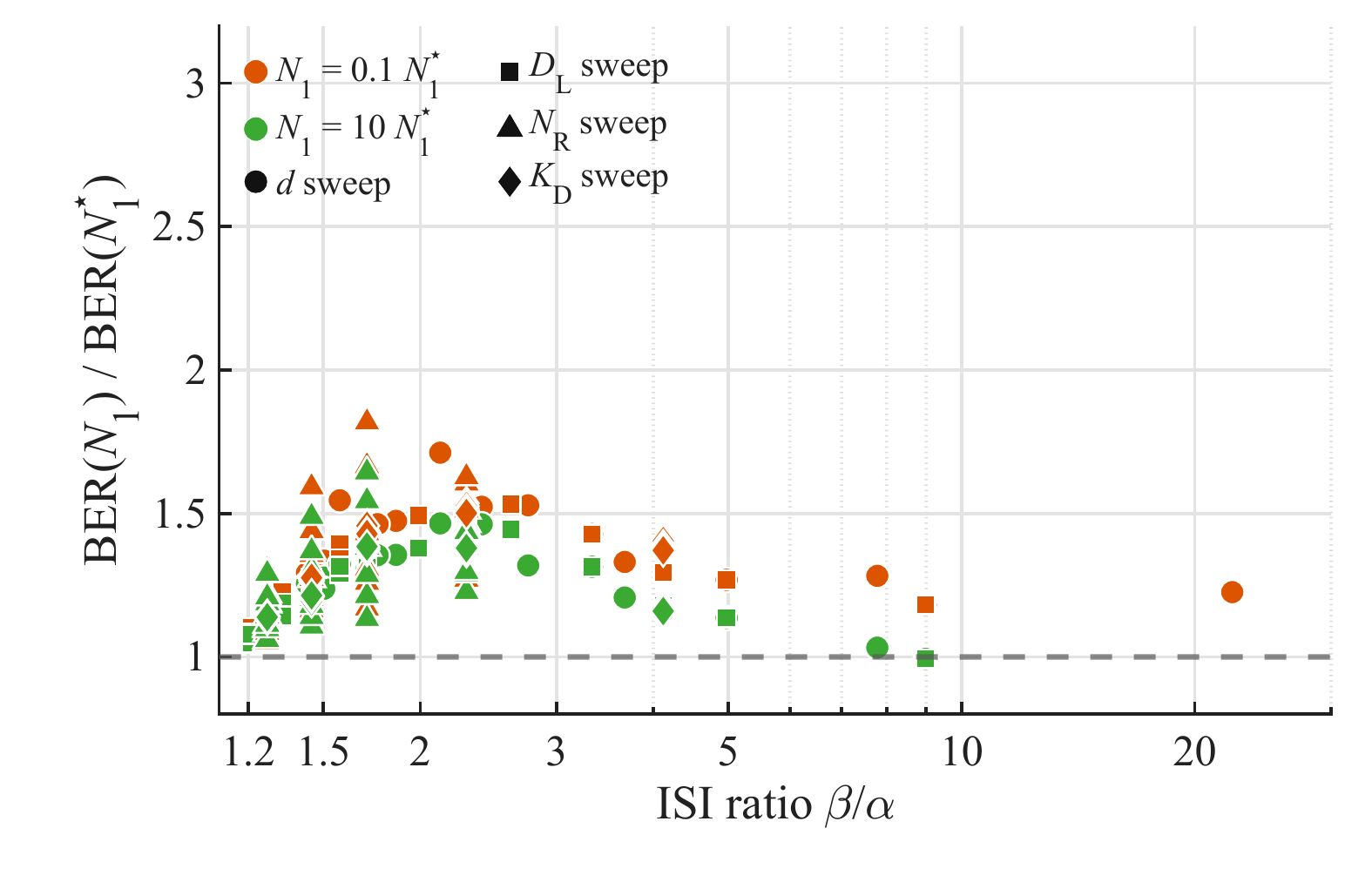}
\caption{Effect of suboptimal release counts. BER inflation factor $\Pe(\None)/\Pe(\Nstar)$ for $\None=0.1\Nstar$ and $10\Nstar$ over all $87$ conditions with BER below $0.45$, versus the ISI ratio (marker shape, sweep; the default configuration appears once per sweep).}
\label{fig:gains}
\end{figure}

Across the $d=10\,\um$ family of Fig.~\ref{fig:base}, releasing one tenth of the predicted count increases the error probability by a factor of roughly $1.3$ to $1.7$ in TD-MC and by up to a factor of two in the particle simulations. Releasing ten times the predicted count costs a comparable factor whenever the ISI ratio is moderate ($\beta/\alpha\le4$) and has a negligible effect only when ISI has essentially vanished, and two orders of magnitude below the prediction the BER exceeds $0.4$ under moderate ISI, close to that of an unusable channel. Fig.~\ref{fig:gains} collects these ratios across all $87$ conditions with BER below $0.45$, where the cost of an order-of-magnitude error clusters between $1.2$ and $1.8$. Since the logarithmic sensitivities of Section~\ref{sec:properties} guarantee that realistic parameter uncertainties displace the operating point by far less than a decade, the closed form removes the dominant, order-of-magnitude uncertainty in the release count at negligible computational cost, leaving a residual penalty of a few percent.

\section{Discussion}
\label{sec:discussion}

\subsection{Design Implications}
Evaluating \eqref{eq:Nstar_closed} requires only design-time quantities such that nothing is measured during operation, no feedback channel is needed, and the receiver remains unchanged, while the logarithmic sensitivities of Section~\ref{sec:properties} ensure that performance degrades only gradually under parameter uncertainty.

The geometric-mean condition $\KD=\sqrt{C_0C_1}$ can thus be enforced from either end of the channel. A receiver able to retune its receptor affinity should match $\KD$ to the prevailing concentration levels, as shown in our previous work~\cite{Kuscu2026AdaptiveReceiver}, whereas a transmitter facing a fixed receptor should scale the levels to match $\KD$, as shown here. The two mechanisms are complementary, and a system able to exercise both gains a family of optimal configurations along which the molecule budget can be traded against receptor engineering effort.

Because repeated-bit decisions rely on interference drift, the comparator's BER exceeds that of coherent detectors with full channel knowledge; our rule optimizes a plausible cell-scale detector. Longer symbols, more receptors (Fig.~\ref{fig:sweeps}(a)), transition-based line coding, and error-correcting codes may reduce the BER further while leaving the optimal release count governed by the same condition.

\subsection{Limitations and Future Work}
Three limitations bound the present claims. First, the analysis rests on the free-space diffusion-based channel without flow, enzymatic degradation, or crowded and bounded geometry. The two-level structure \eqref{eq:levels} survives any linear time-invariant channel, and the rule generalizes verbatim once the corresponding impulse response and its sampled ISI sum are supplied. Degradation in particular shortens the interference tail and admits the same Euler--Maclaurin treatment with an exponentially weighted kernel. Second, the memoryless receptor model assumes kinetics fast relative to the symbol period. The TD-MC layer retains the kinetic lag and shows no visible displacement of the optimum in the studied regime, but markedly slower receptors would correlate consecutive samples and call for a kinetic extension of the rule. Third, the asymptotic ISI coefficient represents an infinitely long transmission, whereas any actual transmission carries less interference, which together with the repeated-bit refinement of Remark~\ref{rem:history} explains why the empirical optimum, when it deviates at all, sits slightly above the prediction. A refinement accounting for both effects is a natural next step, as is estimating the level slopes $\alpha$ and $\beta$ online from receptor statistics, which would turn the static design rule into a fully adaptive transmitter.

\section{Conclusion}
\label{sec:conclusion}
This paper addressed the number of molecules an MC transmitter should release per bit, a fundamental design decision for which no closed-form solution was previously available. When reception occurs through ligand receptors at the receiver and detection through a CSI-free comparison of consecutive bound-receptor counts, the optimal release count follows the closed-form transmission rule $\Nstar=\KD/\sqrt{\alpha\beta}$, which makes the receptor dissociation constant the geometric mean of the two bit-conditioned received concentration levels and which we proved exactly optimal for the underlying memoryless receptor model. Time-domain Monte Carlo sweeps of the channel and receptor parameters, corroborated by particle-based simulations, confirmed that wherever detection is possible at all the empirically optimal release count coincides with the prediction or lies above it by at most a small factor. 

\balance
\bibliographystyle{IEEEtran}
\bibliography{references}

\end{document}